\documentclass{WileyMSP-template}

\begin{document}

\pagestyle{fancy}
\rhead{\includegraphics[width=2.5cm]{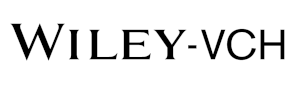}}

\title{The Faraday effect and phase transition \\ in the CH$_3$NH$_3$PbI$_3$ halide perovskite single crystal}

\maketitle


\author{Anastasia A. Shumitskaya}
\author{Vadim O. Kozlov*}
\author{Nikita. I. Selivanov}
\author{Constantinos C. Stoumpos}
\author{Valery S. Zapasskii}
\author{Yury V. Kapitonov}
\author{Ivan I. Ryzhov}

\begin{affiliations}
A. A. Shumitskaya, V. O. Kozlov, Dr. N. I. Selivanov, Prof. C. C. Stoumpos, Dr. V. S. Zapasskii, Dr. Yu. V.
Kapitonov, Dr. I. I. Ryzhov\\
Saint Petersburg State University\\
Ulyanovskaya 1, St. Petersburg 198504, Russia\\
Email Address: vadiim.kozlov@gmail.com
\hfill \break

\end{affiliations}



\begin{abstract}
The spin degree of freedom of charge carriers in halide-perovskite semiconductors can be highly useful for information photonics applications. The Faraday effect is known to be the best indicator of paramagnetism of the material and of the spin-light interaction. In this work, the Faraday effect is demonstrated, for the first time, in a hybrid organic-inorganic halide perovskite MAPbI$_3$ (MA$^+$ = CH$_3$NH$_3^+$). The Faraday rotation and birefringence were measured across the tetragonal-cubic phase transition at 327~K. The Faraday rotation is strongly suppressed below the phase transition temperature due to anisotropy (linear birefringence) of the tetragonal crystal phase. The situation changes drastically above the phase transition temperature, when the crystal becomes optically isotropic. The emerging Faraday rotation obeys the Curie law, demonstrating its population-related paramagnetic nature. This observation opens new prospects for application of these systems and for their investigations using methods of the polarization noise spectroscopy applicable to optically anisotropic materials.

\end{abstract}


\section{Introduction}

Halide perovskites are novel semiconductors with unique combination of properties. Cheap solution synthesis, band gap tunability across the full visible wavelength range and defect tolerance~\cite{kang-hi-defect-tolerance-perovskite17,steirer-defect-tolerance-perovskite16} makes them a promising materials for different optoelectronic applications ranging from photovoltaics~\cite{kojima-organometal-perovskite-photovoltaic09} 
to white light emission~\cite{smith-white-light-perovskite18,stoumpos-ruddlesden-popper-perovskite16} 
and lasing~\cite{shishkin-single-step-laser-perovskite19,murzin-ase-lasing-perovskite20,mamaeva-ultrafast-random-lasing-perovskite22}.
Recent studies have shown spin-optical properties of halide perovskites that make them suitable for applications in spintronics, such as spin-polarized exciton quantum beatings~\cite{odenthal-spin-pol-q-beating-perovskites2017}, spin control of the lasing threshold~\cite{tang-perovskite-spin-laser22}, coherent spin manipulation~\cite{belykh-coherent-spin-dynamics-e-h-perovskite19} and other related magneto-optical phenomena~\cite{shrivastava-polaron-spin-dyn-perovskite20,garcia-arellano-energy-tuning-e-spin-perovskites21,garcia-arellano-unexp-anisotropy-g-factor-perovskite22}, thus opening a way to spinoptronic applications~\cite{ping-spin-optotronic-perovkites18}. In particular, a large Verdet constant and a strong Faraday effect was found in 3D halide perovskites~\cite{sabatini-solution-processed-fr-perovskites20,stamper-magnetic-opt-rotary-MA-perovskite21}.
However, Faraday effect has not been found for the most studied and practically important halide perovskite material -- MAPbI$_3$ (MA$^+$ = CH$_3$NH$_3^+$).

At room temperature, MAPbI$_3$ belongs to the tetragonal crystal  family with I$_{4/mcm}$ space group~\cite{onoda-yamamuro-calorimetric-ir-phase-transition-perovskite90,kawamura-structural-study-perovskite02}. In this connection, the absence of Faraday effect in this material at room temperature is not surprising because it is suppressed by linear birefringence of the crystal. Above 327~K, the crystal undergoes a phase transition to the cubic phase with P$_{m\bar{3}m}$ space group~\cite{kawamura-structural-study-perovskite02}. In this work, we show that Faraday effect in these samples can be correctly measured at elevated temperatures in the cubic phase of the crystal.
We have also carried out complementary polarization measurements around the phase transition and investigated temperature dependence of the Faraday rotation (FR). This quantity obeys the Curie law, which is a proof of the paramagnetic nature of involved spin centers. 




\section{Results}
Figure~\ref{exp-wl}a shows the optical density (OD) spectra of the sample in tetragonal phase at $T=320$~K. This spectrum demonstrates a typical interband absorption of the MAPbI$_3$ semiconductor. Recent study has shown that the binding energy of the exciton in tetragonal phase is of the order of several meV~\cite{Miata_exciton_binding_energy}, so that no excitonic effects are to be expected at these temperatures. The FR was found to be close to zero in the whole studied spectral range in contrast to the MAPbBr$_3$ crystal demonstrating strong FR even at room temperature~\cite{sabatini-solution-processed-fr-perovskites20}.

When the MAPbI$_3$ crystal is heated above $T = 327~K$, it transits into the cubic phase,and its magneto-optical properties change dramatically. Figure~\ref{exp-wl},~b  shows the FR and OD spectra for $T = 333$~K. As is seen, the  OD spectrum remains practically the same (showing that no noticeable changes in the bandgap structure occurs across the phase transition), while a pronounced  FR arises. The FR reaches its maximum at 1.48~eV and is still detectable at 1.33 eV, that was tested in additional experiments. The FR angle, as expected, 
was proportional to the magnetic field $B$ and sample thickness $l$: $\theta = VBl$, where $V$ is the Verdet constant. For $E=1.48$~eV, we obtained $V = 28$~rad$\cdot$T$^{-1}\cdot $m$^{-1}$, which is half as much as for MAPbBr$_3$ and almost an order of magnitude smaller than expected from the Becquerel calculation~\cite{stamper-magnetic-opt-rotary-MA-perovskite21}.

To study behavior of the FR and birefringence across the tetragonal-cubic phase transition, we have tuned the laser to the maximum FR at $E=1.475$~eV.
Figure~\ref{exp-temp} shows temperature dependence of the transmission (a), of the Stokes parameter $S_2$ (b), and of the FR (c). Unlike transmission, behavior of the $S_2$ and FR values strongly changed at the point of the phase transition.  The $S_2$ parameter turned into zero above the phase transition where the sample became optically  isotropic. 
At $T < 327$~K, the sample became birefringent, and strong non-zero $S_2$ signal arose. The FR had the opposite trend: above the phase transition, a strong FR was observed, {while in the low-temperature phase, the FR was getting suppressed and irregular in magnitude.} Complicated behaviour of the $S_2$ and oscillating FR signals, below the phase transition temperature, as we believe, is a result of interplay of temperature dependent linear birefringence of the sample  and its domain structure~\cite{Bari_Ferroelasticity}.\\

Independent measurements of temperature dependence of the FR were carried out in a broader temperature range. Figure~\ref{temp-dep} shows FR for $E=1.475$~eV ($\lambda=840.5$~nm) probe in the $T=303-373$~K range plotted {versus} inverse temperature for magnetic field $B = 3$~mT. The growing FR angle with $1/T$ explicitly indicated presence of a paramagnetic contribution to the Faraday effect. The dashed line, in Fig.~\ref{temp-dep}, shows the fit of the data by the Curie law $\theta = \frac{C}{T} + \theta_p$ with $C=451$~deg$\cdot$K and $\theta_p = -0.68$~deg being the diamagnetic (temperature-independent) contribution to the Faraday effect. The magnitude of the paramagnetic contribution in the studied temperatures range is almost twice larger than diamagnetic. This fact denotes presence of significant unintentional doping. The crystals are known to be characterized by microscopic defects~\cite{murzin-ple-spectrosc-defect-mapi2021} that serve as an intrinsic dopant and affect the electrical and optical properties of the crystal. It should be noted that although direct observation of the spin-related properties by the Faraday effect is not possible in  MAPbI$_3$ at $T < 327$~K, the spin noise spectroscopy, in birefringent phases of the crystal, should be quite applicable~\cite{kozlov-sn-birefringent22}.

\section{Discussion}

Although the Faraday effect is present on a microscopic scale, in MAPbI$_3$, apparently in all phases, its macroscopic manifestation, in birefringent phases,   appears to be suppressed~\cite{nye-phys-properties-crystals85} .
Consider mechanism of this suppression more accurately in terms of the Poincar\'e sphere formalism~\cite{zapasskii-polarized-light-anisotropic-medium99}. 

Polarization state of a fully polarized light can be represented by a Stokes vector $\vec{S} = (S_1, S_2, S_3)$ on the Poincar\'e sphere of unit radius (Fig.~\ref{poincare},a). Components of this vector (the Stokes parameters) describe, respectively, preferential linear polarization of the light in the vertical/horizontal direction, in the direction at +/- 45$^o$ to the vertical, and preferential right/left circular polarization.  In this representation, the changes in the light polarization state are described by characteristic motion  of this vector over the sphere  -- in the equatorial or meridional direction for variation of the polarization plane azimuth or ellipticity, respectively.

The presence of anisotropy can be described, in this representation, by  the {\it anisotropy vector} $\vec{M}$, around which the Stokes vector, like spin in a magnetic field~\cite{zapasskii-polarized-light-anisotropic-medium99}, precesses while propagating through the medium (Fig.~\ref{poincare},a). The vector $\vec{M}$ is aligned along diameter of the  Poincar\'e sphere conneting polarizations of two {\it normal modes}  of the anisotropic medium. Analytically, precession of the Stokes vector in anisotropic medium is described by the following simple equation:

\begin{equation}
\label{eq_der}
\frac{\partial \vec{S}}{\partial x} = [\vec{M} \times \vec{S}],
\end{equation}

where, $x$ is the optical pathlength. This equation formally coincides with the well-known Bloch equation describing precession of spin in a magnetic field, with time substituted for spatial coordinate. This equation clearly shows that spatial frequecy of the Stokes-vector precession (or spatial frequency of polarization beats of polarized light in anisotropic medium) is proportional to magnitude of the anisotropy vector.

Simultaneous action of the Faraday-effect-related circular birefringence (gyrotropy) and linear birefringence can be represented by the anisotropy vector $\vec{M} = (M_{b} \cos \alpha, M_{b} \sin \alpha, M_{FR}) $, where $M_{b}$ is the linear birefringence and $M_{FR} = V B$. Here, $V$ is the Verdet constant, $B$ is the magnetic field strength, and $\alpha$ is the azimuth of the normal-mode  polarization of the medium. In other words, for the medium with both linear and circular birefringence, the anisotropy vector can be represented as a vectorial sum of two vectors $\vec{M_{b}}$ and $\vec{M}_{FR}$. For the initial 'horizontal' polarization state of the light, $\vec{S}_0 = (1,0,0)$, solution of Eq.~(\ref{eq_der}) has the form:

\begin{equation}
\begin{array}{l}
S_1  =  \cos(M_0x) + 
\frac{2 M_b^2}{M_0^2}
\cos^2 \alpha  \sin^2\left(\frac{M_0 x}{2}\right),
\\
S_2  =  
\frac{M_{FR}}{M_0} \sin(M_0x)
+
\frac{M_b^2}{M_0^2}
\sin (2 \alpha) 
\sin^2\left(\frac{M_0 x}{2}\right),
\\
S_3  = 
\frac{2 M_b M_{FR}}{M_0^2} \cos \alpha \sin^2 \left( \frac{M_0 x}{2} \right)
-
\frac{M_b}{M_0}
\sin \alpha
\sin (M_0 x),
\end{array}
\label{stonks}
\end{equation}

where $M_0 = \sqrt{M_b^2 + M_{FR}^2}$.

It is important that, in practice of magneto-optical measurements, the magnetic-field-induced circular birefringence is usually much smaller than linear birefringence of the crystal ($M_{FR} \ll M_b$). Under this condition, as is evident from the picture of precessing spin, the small circular birefringence may only slightly change the direction of the anisotropy vector (controlled by the predominant linear birefringence $\vec{M}_b$) and, thus, cannot substantially affect the precessing Stokes vector. 

The effect of linear birefringence on the measured Faraday rotation is visually illustrated by Fig.~\ref{poincare},b. In the absence of linear birefringence, the anisotropy vector $\vec{M}_{FR}$ is aligned vertically, and the precessing Stokes vector does not leave the equatorial plane of the Poincar\'e sphere. In this case, the polarization-plane rotation angle increases linearly with the light pathlength.  The Stokes vector's tip, in this case, moves over equator of the Poincar\'e sphere.  

In the presence of linear birefringence, the situation dramatically changes. For simplicity, we consider the case when the initial Stokes vector is aligned along the anisotropy vector $\vec{M}_b$ ($\alpha=0$). Figure~\ref{poincare},b shows what happens with trajectory of the Stoke vector's tip with increasing linear birefringence. As the magnitude of the vector  $\vec{M}_b$ increases, the anisotropy vector of the medium  $\vec{M}_0$ is getting closer and closer to the vector  $\vec{M}_b$ and the angle of the Stokes vector precession becomes smaller and smaller (several initial steps of this evolution are shown in  Fig.~\ref{poincare} by numbers (1--5). At $M_{FR} \ll M_b$, which is usually the case, the angle of the precession cone of the Stokes vector becomes extremely small, and the effect of gyrotropy on the Stokes vector appears to be negligible. This picture, as we believe, explains the effect of suppression of the Faraday rotation in the media with linear birefringence.

Quantitatively, the effect of this suppression can be easily evaluated using Eq.~(\ref{stonks}). For small circular birefringence ($|{M_{FR}}|\ll 1$), at $\alpha=0$,  the detected FR angle ($\phi$), which is proportional to the arising $S_2$-component of the Stokes vector, is given by the relation.

\begin{equation}
\begin{array}{l}
\phi \approx
M_{FR} x \,
\frac{\sin(M_b x)}{M_b x}
.
\end{array}
\end{equation} 

The second factor in this equation, which is known to be always smaller than unity, controls suppression of the Faraday rotation in the medium with linear birefringence. This effect is illustrated by Fig.~\ref{poincare},c. As seen from this plot, the detected FR signal rapidly drops with increasing phase retardation between two normal modes of the anisotropic medium. 
The quantity $M_b\cdot x$ may increase either with increasing $\Delta n$, or with increasing thickness of the sample $x$. It is important that, at $M_b\cdot x \gg 1$, absolute value of the FR signal appears to be not only much smaller than what would be observed in isotropic medium, but also dependent on the phase retardation $M_b\cdot x$, and, therefore, practically unpredicted.   At the same time, this figure shows that at small values of the phase retardation ($M_b\cdot x \ll \pi$), the linear birefringence does not preclude measuring the Faraday rotation in a birefringent medium.

\section{Conclusion}

In this work, we succeeded in detecting the Faraday effect in the MAPbI$_3$ halide perovskite single crystal in the cubic phase. 
Previous unsuccessful attempts of such measurements were made at room temperature when the crystal is birefringent  and the Faraday effect proves to be suppressed. Theoretical description of this behavior is presented. 
The Curie-law obeyed contribution to temperature dependence of the Faraday effect  indicates paramagnetic nature of the material and the presence of extrinsic charge carriers in it. The results of this work open a new way for studying the spin subsystem in MAPbI$_3$, including the use of spin noise spectroscopy. In addition, observations of the Faraday effect and birefringence can be used to investigate transformations in a material near the phase transition.


\section{Experimental Section}
The several-millimeter MAPbI$_3$ single crystals were grown by the counterdiffusion-in-gel crystallization method according to the procedure described in~\cite{selivanov-counterdiffusion-in-gel-growth-perovskite22}. 
The magneto-optical setup is shown in Fig.~\ref{setup}. The plane-parallel crystal sample (2) was mounted in the Montana Cryostation vacuum chamber serving as a temperature-controlled thermostat. The sample was illuminated by the intensity-stabilized Ti:sapphire CW laser (Fig.~\ref{setup}). The magnetic field collinear with the light propagation direction was created by the ring coil (3) wound over the sample. The light beam, $\sim$ 4 mm in diameter, was linearly polarized.

The polarization plane azimuth of the transmitted light was adjusted for initial polarimeter balancing by half-wave plate (4) and then split into two channels with 50/50 non-polarizing beamsplitter (5). Degree of linear polarization was measured using  polarizing beamsplitter (6) and powermeters (7). Polarimeter (8) was used to detect rotation of the light polarization plane. This polarimeter consisted of a polarizing beamsplitter and a  balanced photoreceiver PDB450A (Thorlabs) with gain $10^6$ V/A. This differential scheme allows one to suppress the excess intensity noise of the laser light and to achieve the shot-noise limited polarimetric sensitivity~\cite{zapasskii-polarimetry82}. Such polarimetric scheme is widely used in spin noise spectroscopy~\cite{glasenapp-polarimetric-sens13,poltavtsev-sns-single-qw14,ryzhov-sn-explores-local-fields16}.

Solenoid (3) was driven by the AC generator (10) at a frequency of $\sim$1~kHz. The amplitude of the magnetic field at the sample did not exceed $\sim$3 mT. 
The oscillating signal at the output of the balanced detector, proportional to the FR amplitude, was detected using lock-in amplifier (9).   

To measure temperature dependence of the FR and the transmitted light polarization state in the vicinity of phase transition, we tuned the wavelength to the maximum FR at high temperature (well above the phase transition) and allowed the sample to cool down slowly with $\sim$\,5~mK$\cdot\mbox{s}^{-1}$ rate. The normalized debalance signals of detectors (7) and (8) provided, respectively, the DC and AC values of the Stokes parameter $S_2$ ($S_2 = \frac{I_1-I_2}{I_1+I_2}$, where $I_1$ and $I_2$ are the photocurrents of the detectors).

\medskip
\textbf{Acknowledgements} \par 
The sample synthesis, optical experiments (Figs.~\ref{exp-wl},~\ref{exp-temp}) and the theoretical description of the results were supported by Ministry of Science and Higher Education of the Russian Federation Megagrant №075-15-2022-1112 which is higly appreciated. The additional measurements of FR temperature dependence (presented in Fig.~\ref{temp-dep}) were supported by Saint Petersburg State University (Grant No. 94030557). The work was fullfilled using the equipment of Resource Center ``Nanophotonics'' of SPbU Researck Park.

\newpage
\medskip

%
\bibliographystyle{MSP}
\bibliography{faraday-aom}



\vspace{3mm}
\noindent
\begin{minipage}{\linewidth}
\centering
\includegraphics[width=0.9\textwidth]{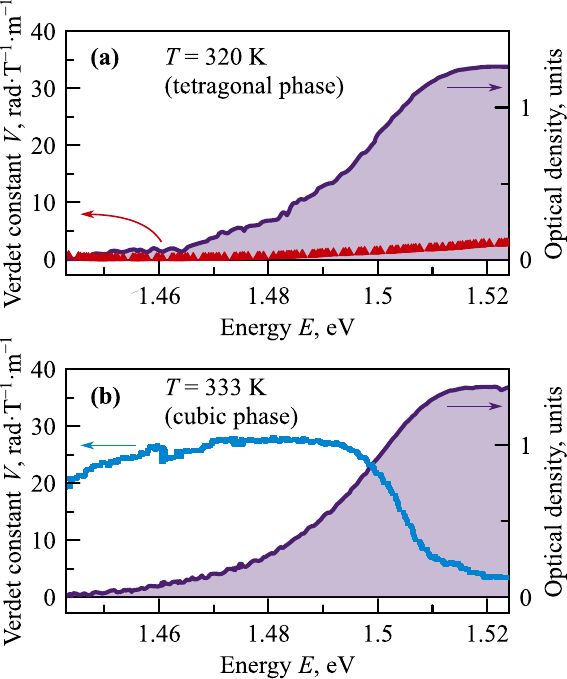}
\captionof{figure}{The Faraday rotation and optical density spectra in the tetragonal (a) and cubic (b) phases of the crystal.
}
\label{exp-wl} 
\end{minipage}
\vspace{3mm}

\vspace{3mm}
\noindent
\begin{minipage}{\linewidth}
\centering
\includegraphics[width=0.9\textwidth]{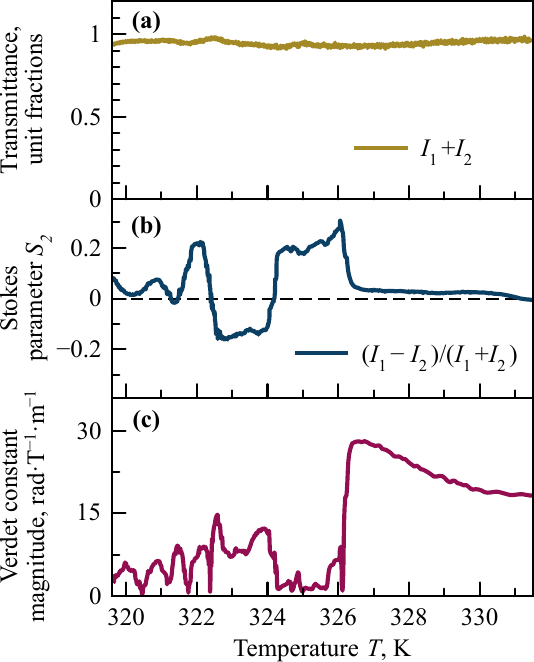}
\captionof{figure}{Temperature dependence of the transmittance (a), Stokes parameter $S_2$ (b) and Verdet constant magnitude measured at $840.5$ nm wavelength. 
}
\label{exp-temp} 
\end{minipage}
\vspace{3mm}

\vspace{3mm}
\noindent
\begin{minipage}{\linewidth}
\centering
\includegraphics[width=\textwidth]{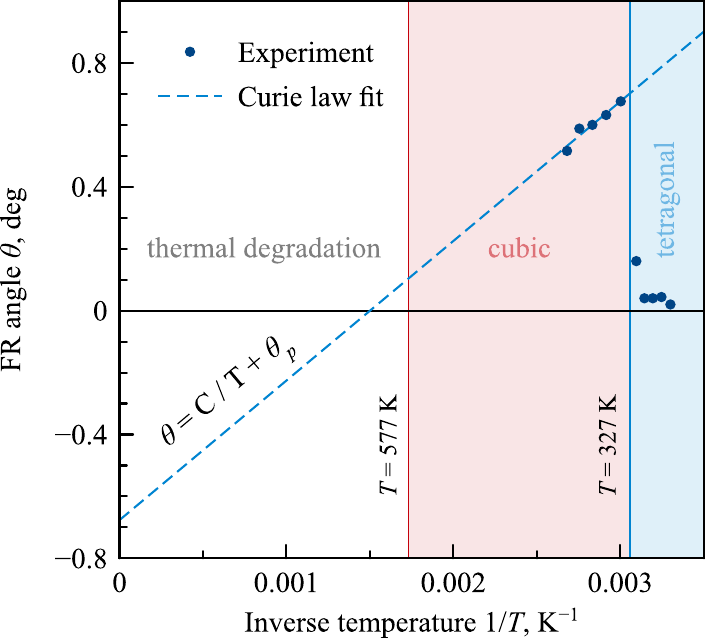}
\captionof{figure}{Temperature dependence of the FR. Dots are the experimental data, dashed line is the Curie-law fit. The areas delimited by temperature values of 327 K and 577 K shows the tetragonal phase (below $T = 327$~K), cubic phase, and the typical thermal degradation temperature for MAPbI$_3$ ($T = 577$~K).
}
\label{temp-dep} 
\end{minipage}
\vspace{3mm}

\vspace{3mm}
\noindent
\begin{minipage}{\linewidth}
\centering
\includegraphics[width=0.9\textwidth]{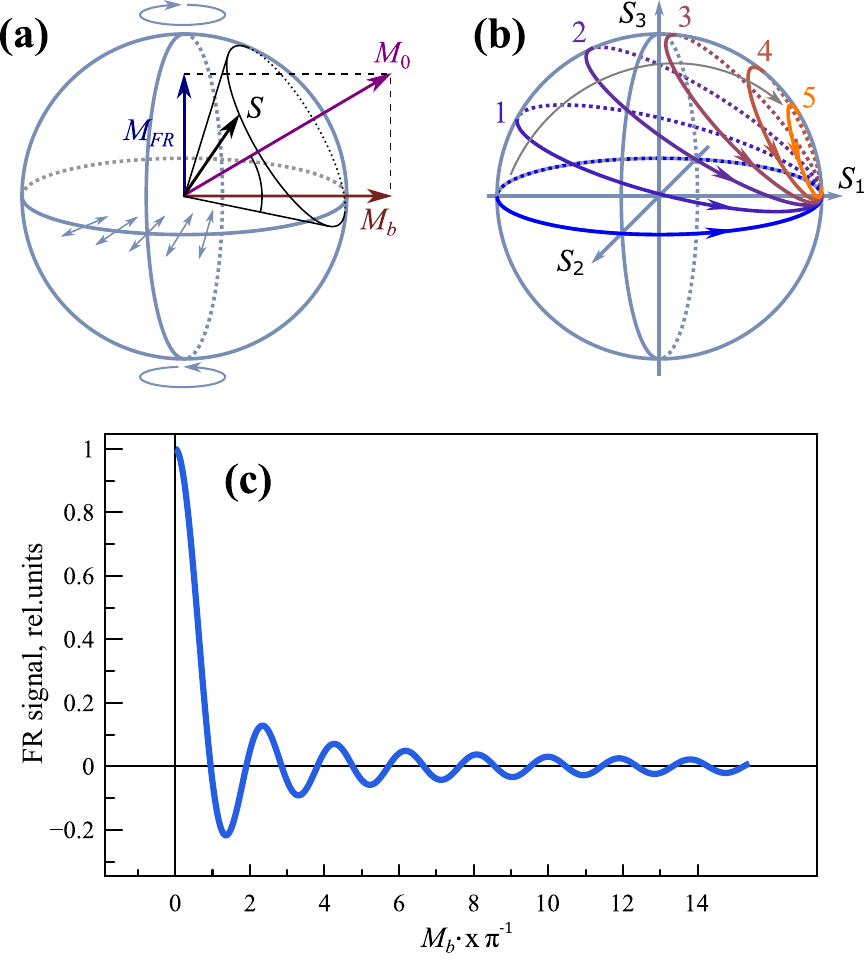}
\captionof{figure}{(a) Representation of polarization evolution of light in 
anisotropic medium by precessing Stokes vector in the Poincar\'e sphere. (b) 
Evolution of trajectories of the Stokes vector precession in a gyrotropic medium with increasing linear birefringence. $M_b/M_{FR}$= 0.3 (1), 0.8 (2), 1.1 (3), 1.8 (4), 2.5 (5).  (c) Effect of suppression of the Faraday rotation in the medium with linear birefringence. The FR in the absence of linear birefringence is taken for unity. The FR is seen to be strongly suppressed when the birefringence-related retardation becomes comparable with the light wavelength.  
}
\label{poincare} 
\end{minipage}
\vspace{3mm}

\vspace{3mm}
\noindent
\begin{minipage}{\linewidth}
\centering
\includegraphics[width=\textwidth]{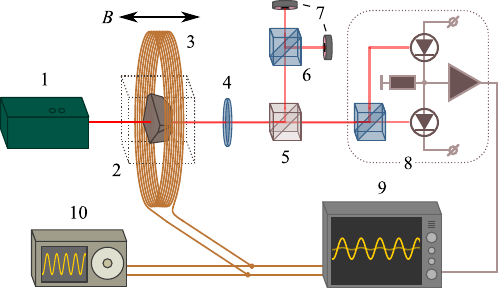}
\captionof{figure}{Experimental setup for FR and light polarization state detection. See notations in the text. 
}
\label{setup} 
\end{minipage}
\vspace{3mm}


\newpage

\begin{figure}
\textbf{Table of Contents}\\
\medskip
  \includegraphics{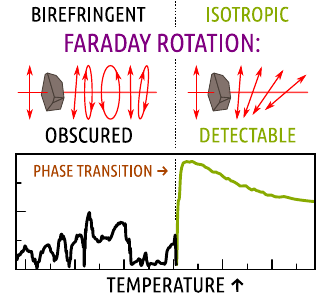}
  \medskip
  \caption*{
  We report on first observation of Faraday rotation (FR) in MAPbI$_3$ single crystal and show its paramagnetic nature. The FR signal is suppressed by arising birefringence at phase transition from cubic to tetragonal phase. We develop a simple model describing that suppresion, and show that the birefringence itself can serve as higly sensitive probe for polarimetric detection of phase transitions.}
\end{figure}

\end{document}